\documentclass[aps,prl,twocolumn,preprintnumbers,amsmath,amssymb,superscriptaddress]{revtex4-1}
\usepackage{graphicx}
\usepackage{subfigure}
\usepackage{epsfig}
\usepackage{dcolumn}
\usepackage{bm}
\usepackage{ulem}
\usepackage{color}
\usepackage{multirow}
\usepackage{slashed}

\def\be{\begin{equation}}       \def\ee{\end{equation}}
\def\bea{\begin{eqnarray}}      \def\eea{\end{eqnarray}}
\def\ba{\begin{array}}
\def\ea{\end{array}}
\def\bnum{\begin{enumerate} }
\def\enum{\end{enumerate}}

\def\nn{\nonumber}

\def\=>{\Rightarrow}
\def\>{\rightarrow}

\def\eye2{Fathbb{I}}

\renewcommand{\>}{\rangle}

\begin{document}

\title{Weyl magnons in pyrochlore antiferromagnets with all-in-all-out orders}

\author{Shao-Kai Jian}
\affiliation{Institute for Advanced Study, Tsinghua University, Beijing 100084, China}

\author{Wenxing Nie}
\email{wenxing.nie@gmail.com}
\affiliation{Center for Theoretical Physics, College of Physical Science and Technology, Sichuan University, Chengdu 610064, China}

\begin{abstract}
We investigate novel topological magnon band crossings of pyrochlore antiferromagnets with all-in-all-out (AIAO) magnetic order. By general symmetry analysis and spin-wave theory, we show that pyrochlore materials with AIAO orders can host Weyl magnons under external magnetic fields or uniaxial strains. Under a small magnetic field, the magnon bands of the pyrochlore with AIAO background can feature two opposite-charged Weyl points, which is the minimal number of Weyl points realizable in quantum materials and has not be experimentally observed so far. We further show that breathing pyrochlores with AIAO orders can exhibit Weyl magnons upon uniaxial strains. These findings apply to any pyrochlore material supporting AIAO orders, irrespective of the forms of interactions. Specifically, we show that the Weyl magnons are robust against direct (positive) Dzyaloshinskii-Moriya interactions. Because of the ubiquitous AIAO orders in pyrochlore magnets including R$_2$Ir$_2$O$_7$, and experimentally achievable external strain and magnetic field, our predictions provide promising arena to witness the Weyl magnons in quantum magnets. 
\end{abstract}
\date{\today}
\maketitle

\textit{Introduction.---}Weyl fermions, which are originally proposed as solutions of massless Dirac equation~\cite{weyl1929}, emerge as linear crossing points \cite{murakami2007, wan2011, balents2011, hasan2015, ding2015, ruan2016a, ruan2016b} of two electronic bands in solid-state materials, the so-called Weyl semimetals. The Weyl points emit Berry flux and behave like monopoles in momentum space, making Weyl points robust again perturbations and leading to many exotic properties. For instance, there are fermi arcs connecting projections of two opposite-charged Weyl points on the surface of Weyl semimetals. It also leads to interesting transport phenomena \cite{nielsen1983, burkov2012, qi2013, son2013, PengYe2013, dai2016}, such as negative magnetoresistivity \cite{son2013}, quantum anomalous Hall effects \cite{burkov2012}, and chiral magnetic effects \cite{burkov2012} etc.

These robust and interesting topological properties trigger the search for topological magnon modes \cite{Romhanyi2015, McClarty2017}, especially the bosonic analog of Weyl points. Weyl (or Dirac) points are discovered in systems with phonons \cite{marquardt2015}, photons \cite{lu2015, wang2016b} as well as magnons \cite{chen2016, mertig2016, wang2016, chen2017, okuma2017, fang2017,owerre2017,balatsky2016}. Like Weyl semimetals, Weyl magnon semimetals exhibit magnon arcs on the surface connecting the projections of opposite-charged Weyl points. Unlike other categories, the Weyl magnon semimetals generically break time-reversal symmetry owing to the presence of magnetic orders. Recently, people have also proposed detections of chiral anomaly of Weyl magnon semimetal by Aharonov--Casher effect \cite{wang2017}. 

Since there exists plenty of distinct magnetic orders in three-dimensional materials, one would naturally ask which material provides the promising platform to realize Weyl magnons. In this letter, by symmetry analysis and model calculations, we show that  under external magnetic fields, pyrochlore antiferromagnets with AIAO orders can host Weyl points. We also show that the breathing pyrochlore \cite{hiroi2014,kimura2013, rau2016} with AIAO order exhibits Weyl magnons upon uniaxial strains. It is the first time that Weyl magnons are shown to exist in materials with AIAO magnetic orders. We emphasize that our symmetry analysis applies to the materials that share the same symmetry and representation with the effective Hamiltonian constructed near $\Gamma$ point, irrespective of the specific forms of complicated interactions, like Dzyaloshinskii-Moriya (DM) interaction \cite{DM1,DM2} as we show later. Since the AIAO orders exist in many heavy transition metal compounds with 4d and 5d elements \cite{wwk2014, kim2016}, such as Nd$_2$Ir$_2$O$_7$ \cite{tomiyasu2012}, Eu$_2$Ir$_2$O$_7$ \cite{sagayama2013, clancy2016}, Cd$_2$Os$_2$O$_7$ \cite{yamaura2012}, and Sm$_2$Ir$_2$O$_7$ \cite{donnerer2016} etc,
and Weyl points can be easily manipulated by external magnetic fields \cite{mertig2016}, our proposals call for experimental detections of Weyl points in pyrochlore antiferromagnets with AIAO magnetic orders.

\begin{figure}	
\subfigure[]{\begin{minipage}[c]{0.25\textwidth}
\centering
\includegraphics[width=2.3cm]{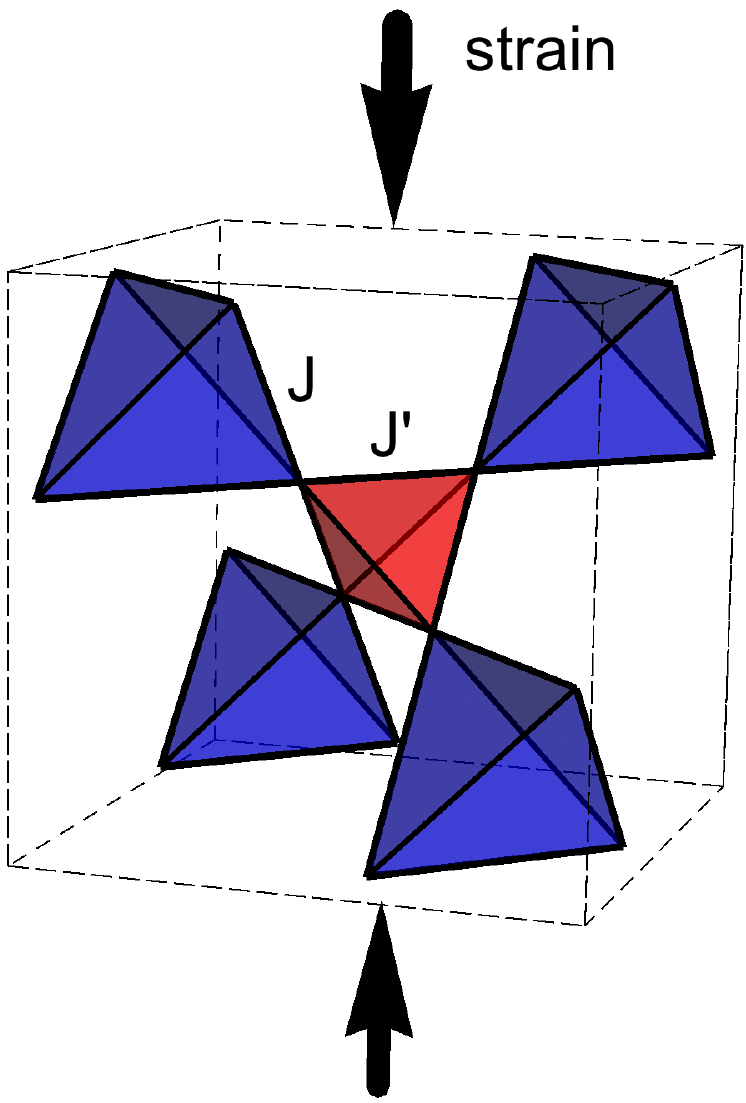}
\end{minipage}}
\subfigure[]{\begin{minipage}[c]{0.2\textwidth}
\centering
\includegraphics[width=4cm]{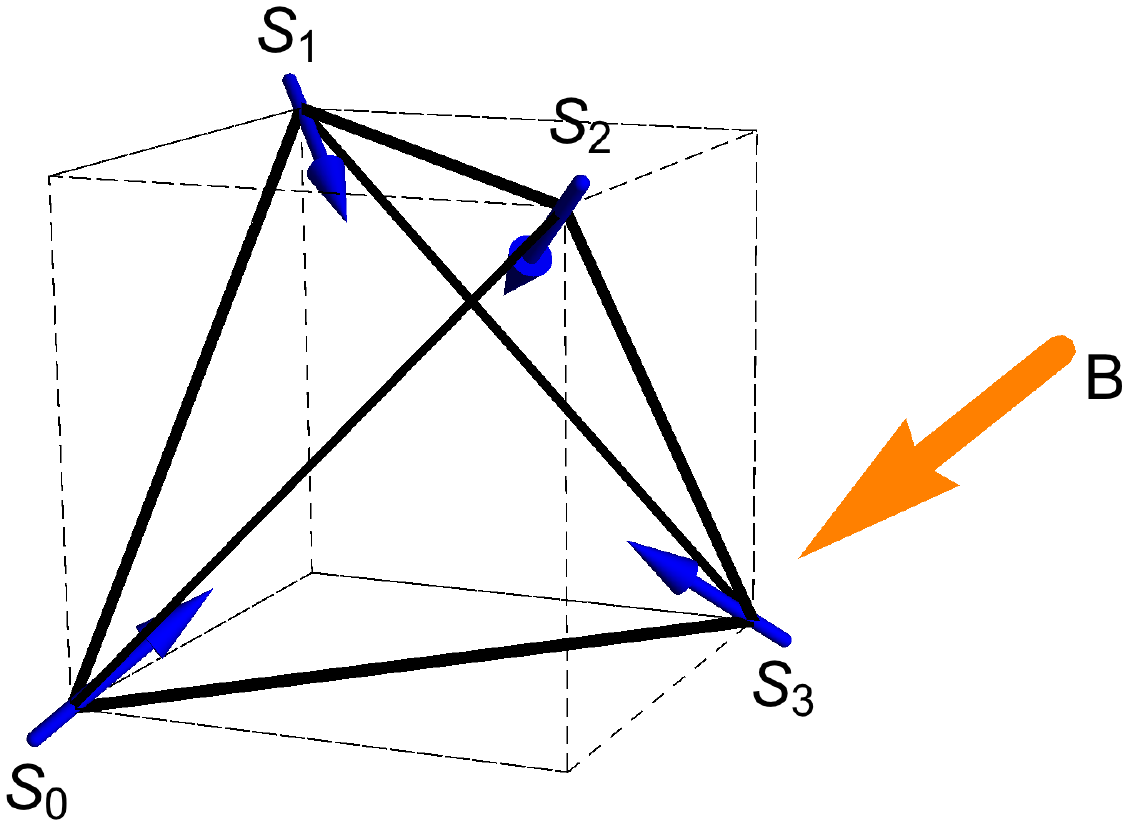}
\end{minipage}}\\
\subfigure[]{\includegraphics[width=3cm]{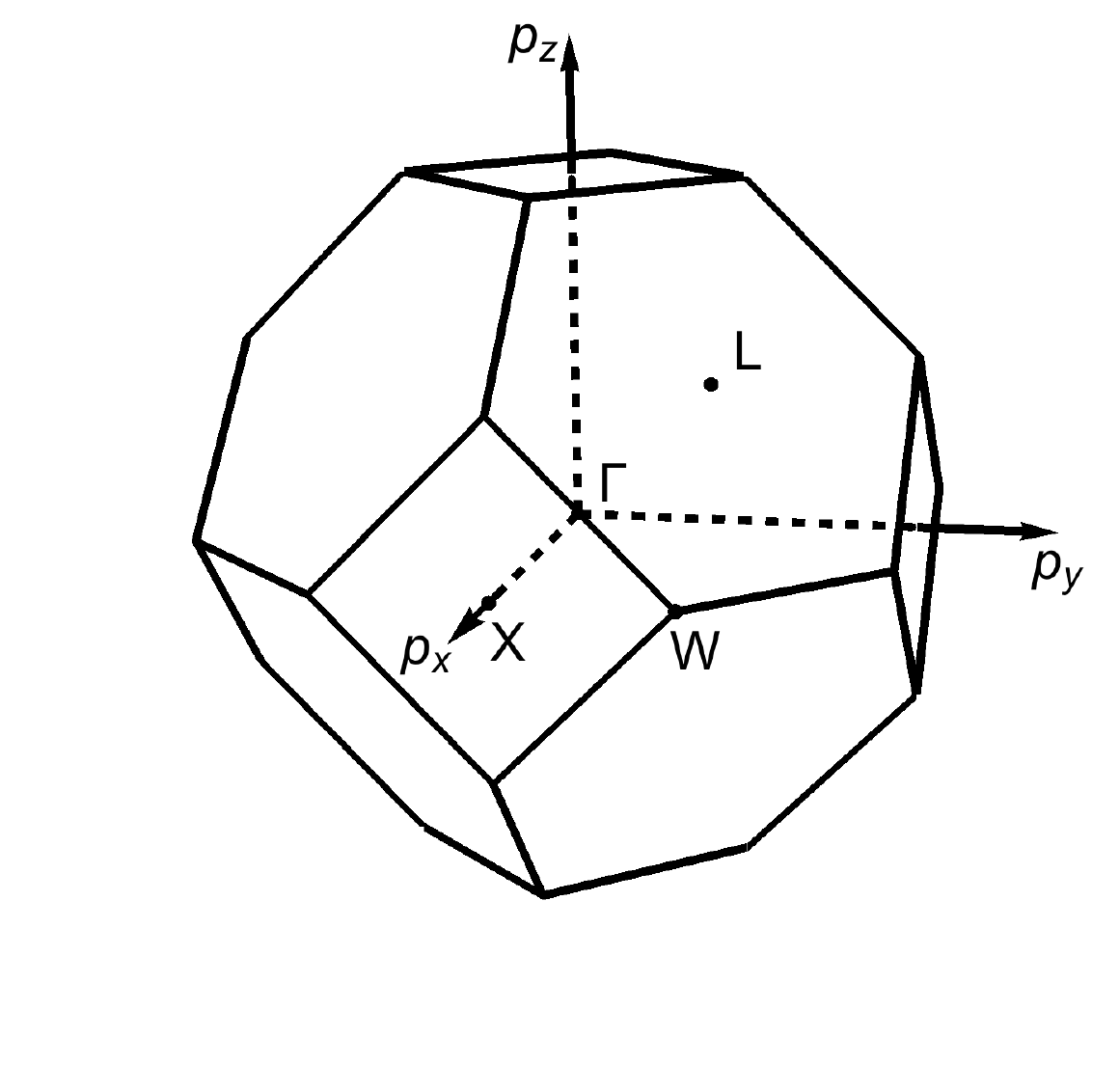}}	~~~		
\subfigure[]{\includegraphics[width=4cm]{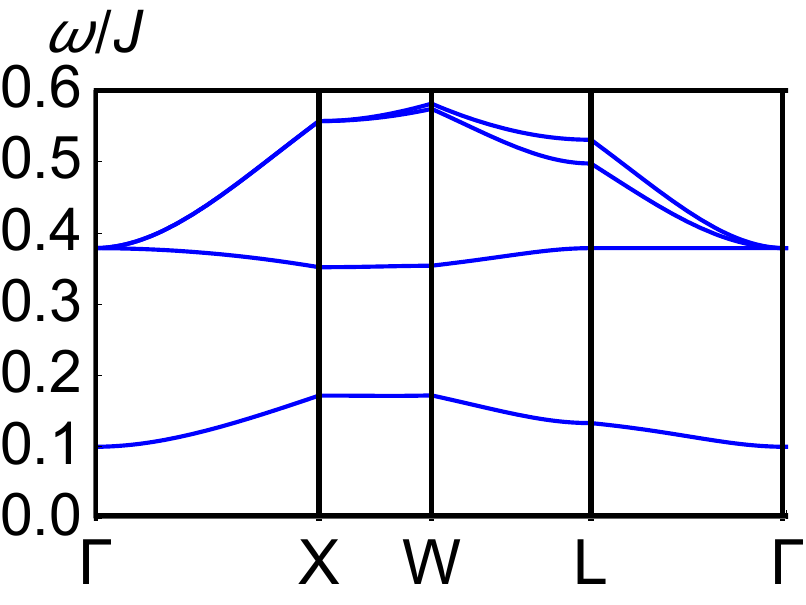}}
\caption{(a) A schematic plot of breathing pyrochlore lattice. The tetrahedron in blue color is up pointing while the tetrahedron in red color is down pointing. The uniaxial strains in $(001)$ direction is shown. (b) A tetrahedra and AIAO order in one cubic, with $\hat z$ directions of local frames. The magnetic fields applied in $(100)$ direction is indicated by the orange arrow. (c) The first Brillouin zone of $fcc$ lattice. The basis of reciprocal lattice are given by $\vec b_1=\frac12 (0,1,1)$, $\vec b_2=\frac12 (1,0,1)$, $\vec b_3=\frac12 (1,1,0)$, where the lattice constant is set to be unit. (d) The magnon bands in pyrochlore lattice with AIAO magnetic order. Note that the four bands split into a triple degeneracy and a singlet at $\Gamma$ point.\label{fig1}}
\end{figure}

\textit{Model and classical phases.---}The point group of pyrochlore lattice is $O_h$, while it is lowered to $T_h$ with the presence of AIAO orders. To model the AIAO magnetic orders, we consider following Hamiltonian \cite{chen2016} on (breathing) pyrochlore lattice,
\bea
	H \!=\! J \! \sum_{\langle ij \rangle \in u} \vec S_i \cdot \vec S_j+ J' \!\sum_{\langle ij \rangle \in d} \vec S_i \cdot \vec S_j+ D \sum_i (\vec S_i \cdot \hat z_i)^2, \label{hamiltonian}
\eea
where $J>0$ and $J'>0$ are antiferromagnetic couplings between nearest neighbors in up-pointing and down-pointing tetrahedra, respectively. As shown in Fig. \ref{fig1}(a), we take $J=J'$ for pyrochlores, and $J\ne J'$ for breathing pyrochlores. The last term in Eq.~\eqref{hamiltonian} describes the local spin anisotropy, where $\hat z_i$ is the local anisotropic direction  pointing to the center of each tetrahedra as indicated by Fig. \ref{fig1}(b). For $D>0$, spin tends to lie in the plane perpendicular to local $\hat z_i$. Thus there is an accidental $U(1)$ degeneracy of classical orders, which is found to be broken by quantum disorder, leading to the existence of  Weyl magnons \cite{chen2016}. 

Here we consider an easy-axis spin anisotropy $D<0$. It is easy to show that the classical ground state is AIAO ordered.
For $D<0$, the third term is maximally satisfied if $\vec S_i$ aligns or anti-aligns in local $\hat z_i$-directions. There are totally sixteen configurations with two AIAO configurations among them. Only AIAO configuration satisfies the local constraint $\sum_{i \in u(d)} \vec S_i=0$ from the first two terms to optimize the exchange interactions simultaneously. 

\textit{Symmetry analysis.---}A key observation of the magnon bands is that the states at $\Gamma$ point are triply degenerate as shown in Fig. \ref{fig1}(d), a spin-wave spectrum of Eq.~\eqref{hamiltonian} with $D<0$. Thus these degenerate states form a $T_g$ representations of $T_h$ group in the presence of AIAO order, distinct from other magnetic orders in pyrochlore, e.g. the easy-plane orders \cite{chen2016}, where the $\Gamma$ point is at most doubly degenerate. 

We first consider pyrochlore lattice with AIAO magnetic order, which preserves $T_h$ symmetry,
and will show the presence of two Weyl points under external magnetic fields. Given that the states at $\Gamma$ point form three-dimensional $T_g$ representations of $T_h$ group, we can construct an effective $k\cdot p$ theory near the $\Gamma$ point. The Hamiltonian up to quadratic order in momentum space is constructed in Ref. \cite{luttinger1956, supp} and is given by
\bea
	 \mathcal{H}_{T}(\vec{p}) &=&  \alpha_1 |\vec{p}|^2 + \alpha_2 \sum_i p_i^2 L_i^2 \nn\\
	 && + [p_x p_y( \alpha_3 \{ L_x, L_y \} + \alpha_4 L_z)+ c.p.], \label{h0}
	 \label{eq.effectiveH}
\eea
where $L_i (i=1,2,3)$ is a 3$\times$3 matrix \cite{supp} and $\alpha_j (j=1,...,4)$ is a constant which characterizes the dispersion around $\Gamma$ point. $\{\}$ denotes anticommutator and $c.p.$ means cyclic permutations. Note that $\alpha_4$ term breaks time-reversal symmetry. The subscript $T$ indicates $T_h$ group.

\begin{figure}
\subfigure[]{\includegraphics[width=3.cm]{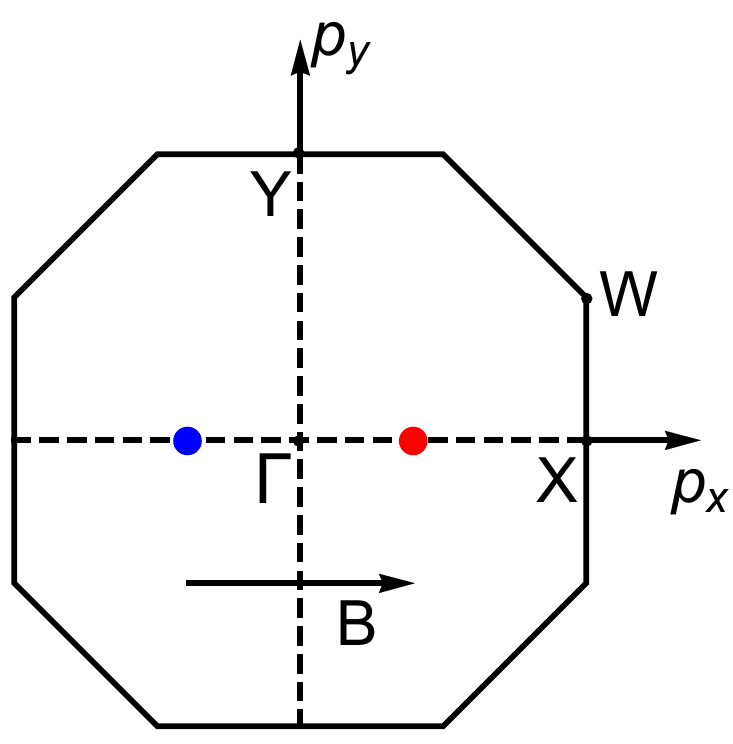}}~~~
\subfigure[]{\includegraphics[width=4.4cm]{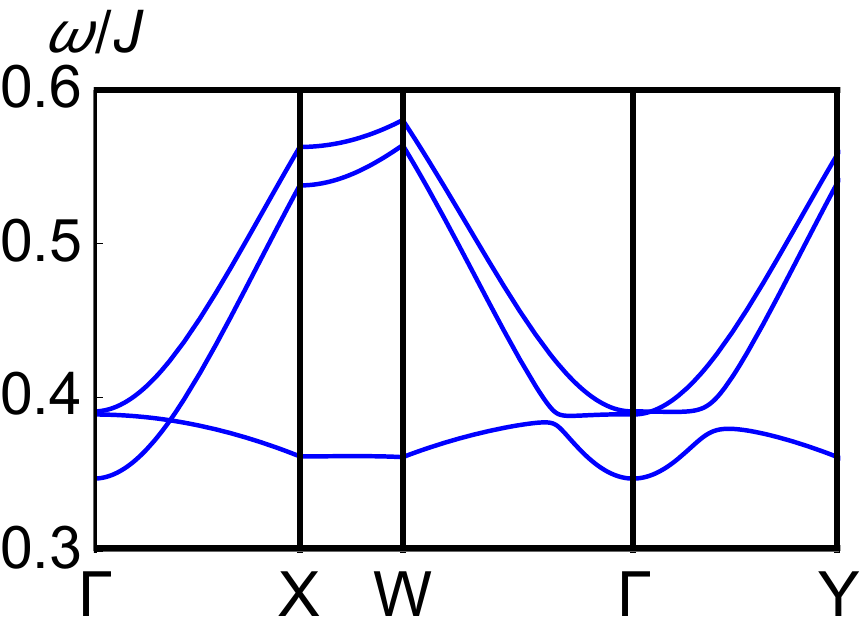}}\\
\subfigure[]{\includegraphics[width=3.cm]{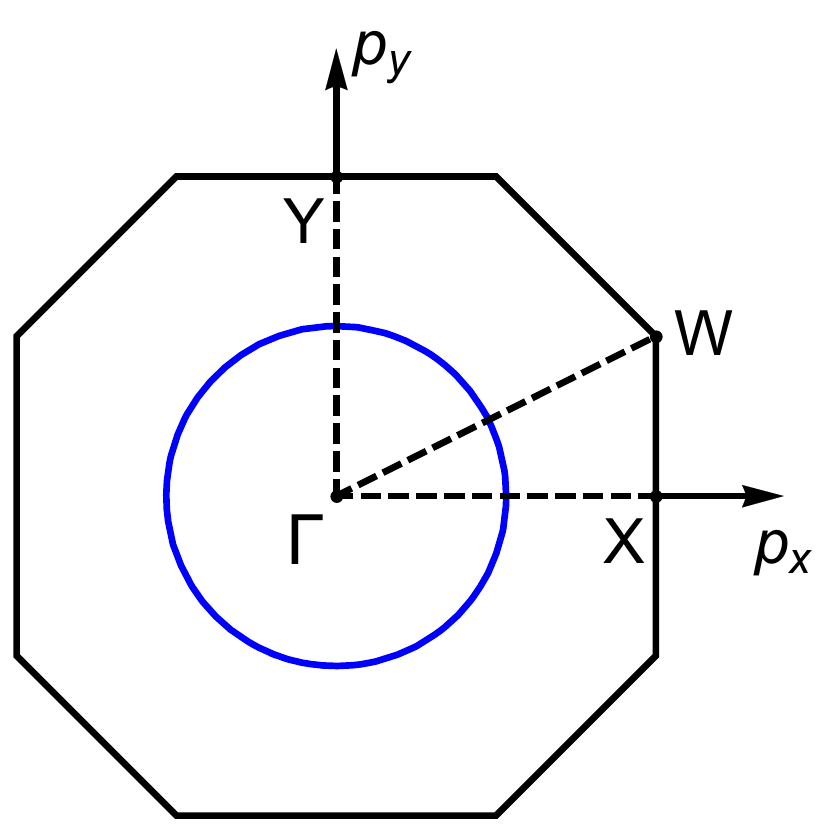}}~~~
\subfigure[]{\includegraphics[width=4.4cm]{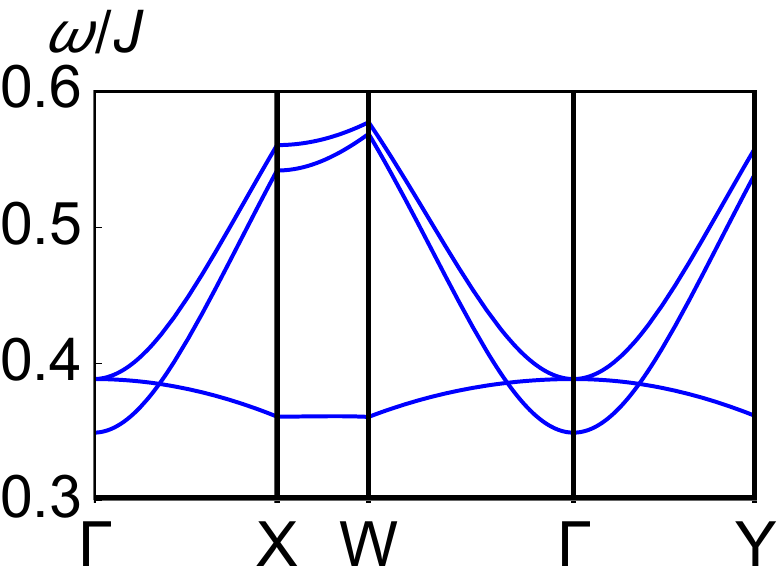}}
\caption{(a) A schematic plot of two Weyl points at $p_x$ axis. Red and blue colors denote opposite monopole charges. The applied magnetic field is shown. (b) The magnon bands along the path $\Gamma$-$X$-$W$-$\Gamma$-$Y$ when magnetic fields are applied, with $S=1/2, J'=J, D=-0.2J, B=0.05 J$. The band-crossing point is along $\Gamma$-$X$.  (c) A schematic plot to show a nodal line emerges in the pyrochlores upon a uniaxial strain along $(001)$ direction. (d) The magnon bands along  $\Gamma$-$X$-$W$-$\Gamma$-$Y$ path with $S=1/2, J'=J, D=-0.2J, \gamma=2\%, B_z=0.05 J$. It clearly shows the nodal-line crossings.}
\label{fig2}
\end{figure}

There is a double degeneracy along each axes, namely, $(100)$, $(010)$ and $(001)$, as shown in Fig. \ref{fig1}(d), protected by $C_2$ rotational and $\sigma_h$ horizontal reflection symmetries. A simple strategy to get Weyl magnons is to break these symmetries, especially to split these two-fold degenerate bands. 
An experimentally accessible way to lower the symmetry is applying an external magnetic field. 
Accordingly, 
we introduce a Zeeman term  $\mathcal{H}_{Z}=\vec\beta_Z \cdot \vec L$ into Eq.~\eqref{eq.effectiveH}, where $\vec \beta_Z$ is proportional to applied magnetic field. The subscript $Z$ indicates Zeeman effect.
For simplicity, we assume the magnetic field is along $(100)$ direction as indicated by orange arrow in Fig. \ref{fig1}(b). The Zeeman term is simplified as $\mathcal{H}_{Z}=\beta^x_Z L_x$. Then the symmetry group is lowered from $T_h$ to $C_{2h}$. The double degeneracy along $(010)$ and $(001)$ directions is split entirely; while the degeneracy along $(100)$ direction is split except two crossings at $(\pm Q_1,0,0)$, where $Q_1=\sqrt{|\beta^x_Z/\alpha_2|}$.
Thus we get a minimal Weyl magnon band structure from $\mathcal{H}_T+\mathcal{H}_Z$ with only two Weyl points as shown in Fig. \ref{fig2}(a). Note that the Weyl points are locked at $p_x$ axis owing to $C_2$ rotational symmetry. In general, the magnetic field can be applied in arbitrary direction. Two Weyl points still survive but will be shifted away from $p_x=0$ axis as long as the magnetic field is small enough. Note that 
the Weyl magnons also appear in breathing pyrochlore with AIAO orders when magnetic fields are applied.

Another convenient way to lower the symmetry is applying strains upon the materials \cite{ruan2016a, ruan2016b, wang2016}. For instance, uniaxial strain along $(001)$ direction lowers the symmetry to $D_{2h}$ group, giving rise the following term,
\bea
	\mathcal{H}_{\text{strain}}= \beta_{\text{strain}}(2L_z^2-L_x^2-L_y^2),
	\label{H_strain}
\eea
where $\beta_\text{strain}$ is a real constant corresponding to the strength of the strain. In the presence of $\mathcal{H}_\text{strain}$, there emerges a nodal line in $p_z=0$ plane, if $\alpha_2\cdot\beta_{\text{strain}}>0$, as shown in Fig. \ref{fig2}(c) (also see Supplemental Materials \cite{supp} for details).
This nodal line is protected by $\sigma_h$ horizontal reflection symmetry \cite{supp}.
It may lead to flat magnon surface states \cite{mertig2017, hu2017}. 

\begin{figure}[t]
\subfigure[]{\includegraphics[width=3.cm]{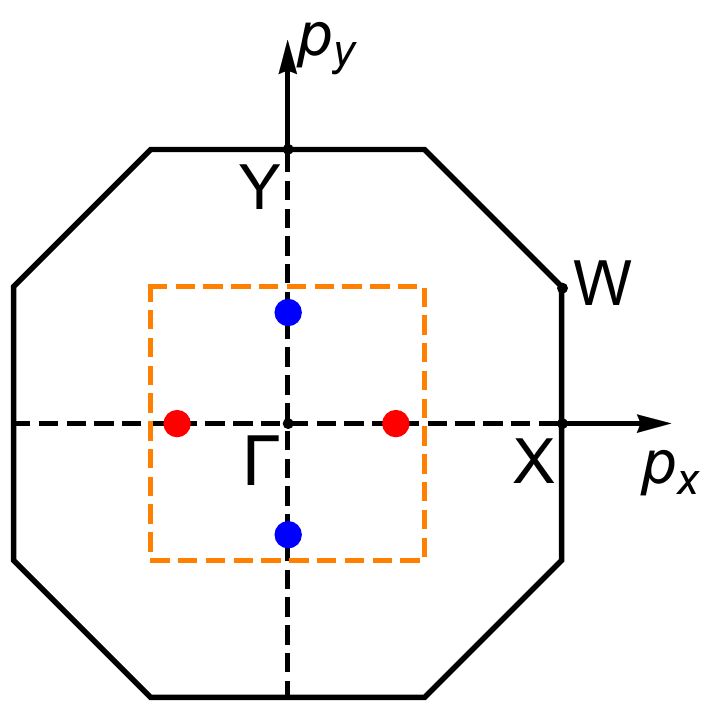}}~~~
\subfigure[]{\includegraphics[width=4.cm]{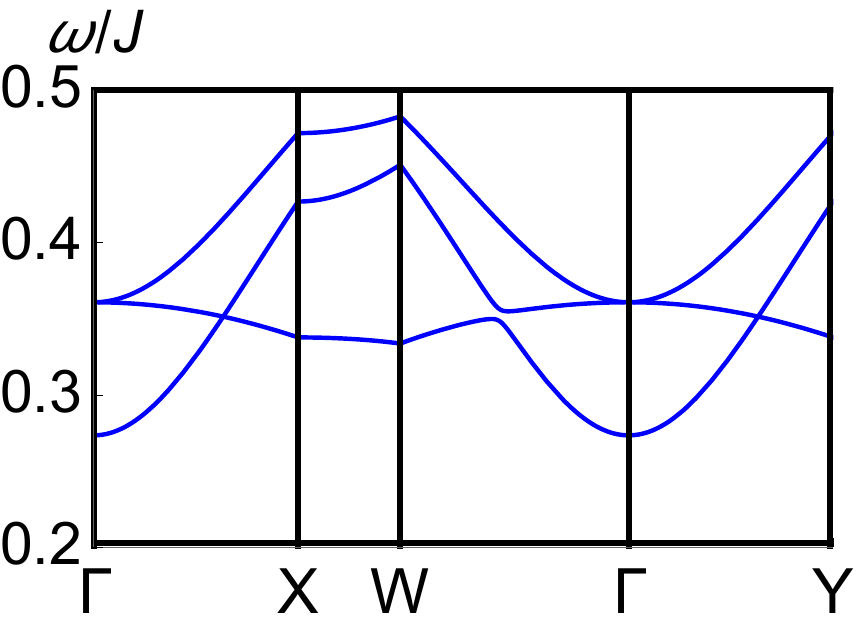}} \\
\subfigure[]{\includegraphics[width=3.cm]{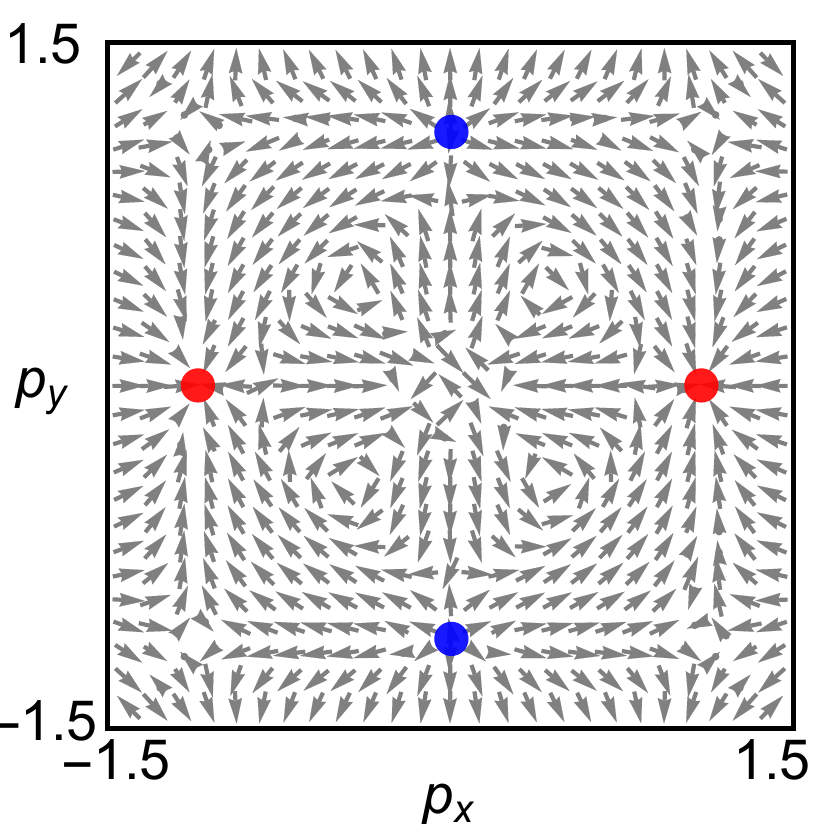}}~~~
\subfigure[]{\includegraphics[width=3.cm]{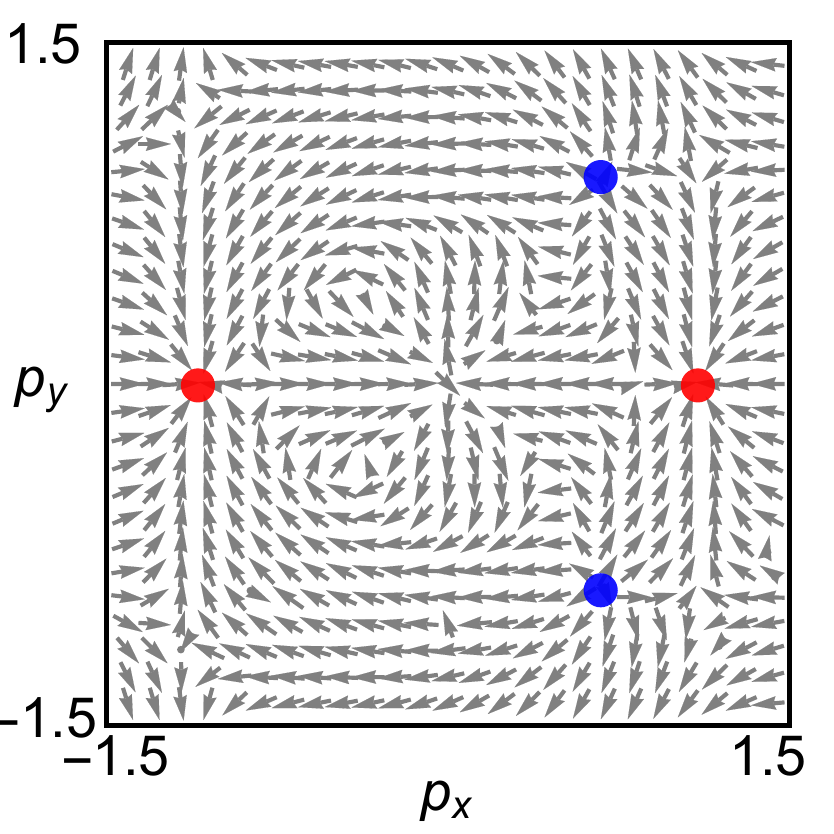}}
\caption{Four Weyl points emerge in breathing pyrochlore upon uniaxial strains along (001) direction. (a) A schematic plot of four Weyl points. Different colors denote opposite monopole charges in $p_z=0$ plane. (b) The magnon bands along the path $\Gamma$-$X$-$W$-$\Gamma$-$Y$, with $S=1/2, J'=0.6J, D=-0.2J, \gamma=2\%$, showing the Weyl points are located at $p_x$ and $p_y$ axis. Berry curvatures of the four Weyl points within the orange frame in (a) are plotted in (c) and (d). (c) The projections of the directions of Berry curvature to $p_z=0$ plane for $p_{x,y} \in(-1.5,1.5)$. It shows the four Weyl points with monopole charge 1 (blue color) and $-1$ (red color). (d) The projections of  directions of Berry curvature to $p_z=0$ plane with applied magnetic field $B_x=0.005J$.}
\label{fig3}
\end{figure}

Now, we consider the breathing pyrochlore lattice, in which  $O_h$ group of pyrochlore lattice is lowered into $T_d$ group. (Note that AIAO order in breathing pyrochlore preserves $T$ group). Follow the same strategy, we add $(001)$-directional strains to lower the symmetry from $T$ group down to $D_{2}$ group. 
Accordingly, we find besides Eq.~\eqref{H_strain}, the following term is also allowed by symmetry,
\bea
	\mathcal{H}_{D}(\vec p)= \beta_D (p_x L_x- p_y L_y),
\eea
where $\beta_D$ is a real constant corresponding to the strength of strains. The subscript $D$ indicates $D_2$ group. After adding 
$\mathcal{H}_\text{strain}+ \mathcal{H}_D$ to $\mathcal{H}_T$, the nodal line splits into four Weyl points located at the points $(\pm Q_2,0,0)$ and $(0,\pm Q_2,0)$, since the the horizontal reflection symmetry $\sigma_h$ is broken, where $Q_2$ is given in \cite{supp}. A schematic plot of these Weyl points is shown in Fig. \ref{fig3}(a).

\textit{Weyl magnons in pyrochlores under a magnetic field.---}By symmetry analysis, with the background of AIAO magnetic order, the Weyl magnons and nodal-line magnons \cite{mertig2017, hu2017} appear under external magnetic fields or uniaxial strains. Now we use linear spin-wave theory to show the emergence of Weyl magnons explicitly. The spin operator can be expressed in terms of the Holstein--Primakoff bosons  $\vec S_\mu \cdot \hat z_\mu = S- a_\mu^\dag a_\mu$, $\vec S_\mu \cdot \hat x_\mu= \sqrt{2S} (a_\mu+ a_\mu^\dag)/2 $, and $\vec S_\mu \cdot \hat y_\mu= \sqrt{2S} (a_\mu- a_\mu^\dag)/(2i) $, where $a_\mu$ $(a_\mu^\dag)$ is annihilation (creation) operator of Holstein--Primakoff boson at $\mu$th sublattice, and the local frames at each sublattice are listed in Appendix. The spin-wave Hamiltonian up to quadratic order is given by
\bea
H= \sum_{\vec p} \sum_{\mu\nu} [a_{\vec p, \mu}^\dag A_{\mu\nu} (\vec p) a_{\vec p, \nu}+ (a_{\vec p, \mu} B_{\mu\nu} (\vec p) a_{-\vec p, \nu}+ H.c.) ],  \nn\\
\eea
 where 
$A_{\mu\nu}(\vec p) \!=\! S[\delta_{\mu\nu} (J\!+\!J'\!-\!2D) \!-\! \frac13 (1\!-\!\delta_{\mu\nu})(J \!+\!J' e^{i (p_\mu-p_\nu)})]$, and $B_{\mu\nu}(\vec p)=S(1-\delta_{\mu\nu}) e^{i \phi_{\mu\nu}}(J+J'e^{-i(p_\mu-p_\nu)})$ with $\phi_{01}=\phi_{23}=-\frac{\pi}{3}, \phi_{02}=\phi_{13}=\frac{\pi}{3}$ and $\phi_{03}=\phi_{12}=\pi$, are 4$\times$4 matrices in sublattice space. And $p_\mu\equiv \vec p \cdot \vec b_\mu$, with $\vec b_0=(0,0,0)$ and $\vec b_{1,2,3}$ defined in the caption of Fig. \ref{fig1}(c). The spin-wave spectrum is shown in Fig. \ref{fig1}(d), where the coupling constants are chosen to be $J=J', D=-0.2J, S=1/2$. A gap between the first and the upper three bands at $\Gamma$  is obvious.

Consider applying a magnetic field, i.e., $\delta H_Z= \vec B \cdot \sum_i \vec S_i$, where $\vec B$ is the magnetic field (we have absorbed the coupling into magnetic field, thus $\vec B$ has the dimension of energy), 
in the linear spin-wave regime (i.e., the magnetic field are small enough), we get $\delta A^Z_{\mu\nu} = \sum_{j=x,y,z} B_{j} \delta A^Z_{j,\mu\nu}$, where 
$\delta A^Z_{x}= \text{diag}(-1,-1,1,1)/\sqrt{3}$, 
$\delta A^Z_{y}= \text{diag}(-1,1,-1,1)/\sqrt{3}$ and 
$\delta A^Z_{z}= \text{diag}(-1,1,1,-1)/\sqrt{3}$.
To be specific, we consider a small magnetic field in $(100)$ direction, i.e., $\vec B=(B,0,0)$.  The spectrum of $H+\delta H_Z$ is plotted in Fig. \ref{fig2}(b), showing a Weyl point along $\Gamma$-$X$ line. The two Weyl points are restricted in $p_x$ axis due to $C_2$ rotational symmetry along $(100)$ direction. The magnetic field can be along any direction and it shift the positions of the two Weyl points, realizing a tunable  Weyl points by external magnetic fields \cite{mertig2016}.

\textit{Nodal-line magnons in strained pyrochlores.---}Upon (001) uniaxial strains, the antiferromagnetic exchange interactions within an up-pointing (down-pointing) tetrahedra become inequivalent, i.e., the two bonds lying in $xy$ plane are distinct from the rest four bonds. 
The coupling strengths become $J \rightarrow (1 \pm \gamma) J$, 
where $\gamma$ characterizes the effect of the strain. 
The perturbation introduced by stains is captured by 
$ \delta H_\text{strain} = \sum_{\vec r} [\gamma J (\vec S_{\vec r, 0} \cdot \vec S_{\vec r, 3}+ \vec S_{\vec r, 1} \cdot \vec S_{\vec r, 2})  
 -\gamma J (\vec S_{\vec r, 0} \cdot \vec S_{\vec r, 1} + \vec S_{\vec r, 0} \cdot \vec S_{\vec r, 2}+ \vec S_{\vec r, 3} \cdot \vec S_{\vec r, 1}+ \vec S_{\vec r, 3} \cdot \vec S_{\vec r, 2})$, where the summation is over all unit cells. It leads to $\delta A^\text{strain}_{\mu\mu}=\gamma A_{\mu\mu}/3$, and $\delta X^\text{strain}_{\mu\nu}= \gamma X_{\mu\nu}$ for $\mu+\nu=3$, $\delta X^\text{strain}_{\mu\nu}= -\gamma X_{\mu\nu} $ for $\mu\ne \nu, \mu+\nu\ne3$, where $X=A,B$. The spectrum of magnons with a small tensile strain $\gamma=2\%$ is shown in Fig. \ref{fig2}(d). The crossing points in $\Gamma$-$X$, $\Gamma$-$Y$ and $\Gamma$-$W$ lines show the evidence of the predicted nodal line in $p_z=0$ plane, which is protected by $\sigma_h$ symmetry.

The applied strains will generically modify the background AIAO order, since the constraint $\sum_{i \in u(d)} \vec S_i=0$ can no longer be satisfied. However, if the strain is much smaller than $J$, $J'$ and $D$ ($\gamma= 2\%$ in our case), we can neglect the titling of the background from AIAO order. 
Note that the local frames are also adjusted corresponding to the strains, but we can neglect this effect since it only shifts the positions of the nodal lines.

\textit{Weyl magnons in strained breathing pyrochlore.---}The presence of two distinct tetrahedra in breathing pyrochlore, where  $J\ne J'$, lowers the symmetry from $O_h$ group to $T_d$ group. 
The symmetry is further lowered by the applied strain in (001) direction to $D_{2}$ group. The effect of stains is captured by replacing $J, J'$ by $(1\pm\gamma)J$ and $(1\pm \gamma')J'$, respectively, leading to
$\delta H'_\text{strain}= \sum_{\vec r \in u} [\gamma J (\vec S_{\vec r, 0} \cdot \vec S_{\vec r, 3}+ \vec S_{\vec r, 1} \cdot \vec S_{\vec r, 2})  -\gamma J (\vec S_{\vec r, 0} \cdot \vec S_{\vec r, 1} + \vec S_{\vec r, 0} \cdot \vec S_{\vec r, 2}+ \vec S_{\vec r, 3} \cdot \vec S_{\vec r, 1}+ \vec S_{\vec r, 3} \cdot \vec S_{\vec r, 2})  + \sum_{\vec r \in d} [\gamma' J' (\vec S_{\vec r, 0} \cdot \vec S_{\vec r, 3}+ \vec S_{\vec r, 1} \cdot \vec S_{\vec r, 2})  -\gamma' J' (\vec S_{\vec r, 0} \cdot \vec S_{\vec r, 1} + \vec S_{\vec r, 0} \cdot \vec S_{\vec r, 2}+ \vec S_{\vec r, 3} \cdot \vec S_{\vec r, 1}+ \vec S_{\vec r, 3} \cdot \vec S_{\vec r, 2})$. 
As predicted by the symmetry analysis, four Weyl points emerge in the $p_z=0$ plane in the spectrum of $H+\delta H'_\text{strain}$ as shown in Fig. \ref{fig3}(b). Actually they are located at $p_x$ and $p_y$ axes due to $C_2$ rotational symmetry. The projections of Berry curvature to $p_z=0$ plane are plotted in Fig. \ref{fig3}(c), where four Weyl points with monopole charge 1 (blue) and $-1$ (red) are clearly shown.

One can again apply magnetic fields to the system, and consequently tune the positions of Weyl points in the reciprocal space \cite{mertig2016}. To explicitly show an example, we apply a small magnetic field $B=0.005J$ in $(100)$ direction. The shifts of Weyl points are shown in Fig. \ref{fig3}(d) indicated by the Berry curvatures. Owing to the $C_2$ symmetry along $(100)$ direction, two Weyl points (red) are still locked in $p_x$ axis.  
Keeping increasing the magnetic fields, two Weyl points with monopole charge 1 (blue) will hit one of the Weyl points with monopole charge $-1$ (red) and then these three Weyl points merge as a single Weyl point with monopole charge $-1$. But there will emerge another four Weyl points \cite{supp}.

\textit{Weyl magnons in the presence of DM interactions.---} Here, we emphasize that the results from the symmetry analysis is applicable not only to the representative model Hamiltonian in Eq.~\eqref{hamiltonian}, but also to all the magnetic materials that satisfy $T_d$ symmetry and whose magnon bands fulfill three-dimensional $T_g$ representation near $\Gamma$ point. Especially, the Weyl magnon can be robust in the presence of the DM interactions which are important in those strongly correlated materials with spin-orbit coupling. To demonstrate that, we add DM interaction as an example,
\bea
	H \!=\!  \! \sum_{\langle ij \rangle} (J\vec S_i \cdot \vec S_j + \vec{D}_{ij} \cdot \vec S_i \times \vec S_j )+ D \sum_i (\vec S_i \cdot \hat z_i)^2,
\label{hamiltonian_DM}
\eea
where DM vector $\vec D_{ij}$ is dictated by $T_d$ group \cite{supp} whose magnitude is given by $|\vec D|=DM$. 

\begin{figure}
\subfigure[]{\includegraphics[width=4.cm]{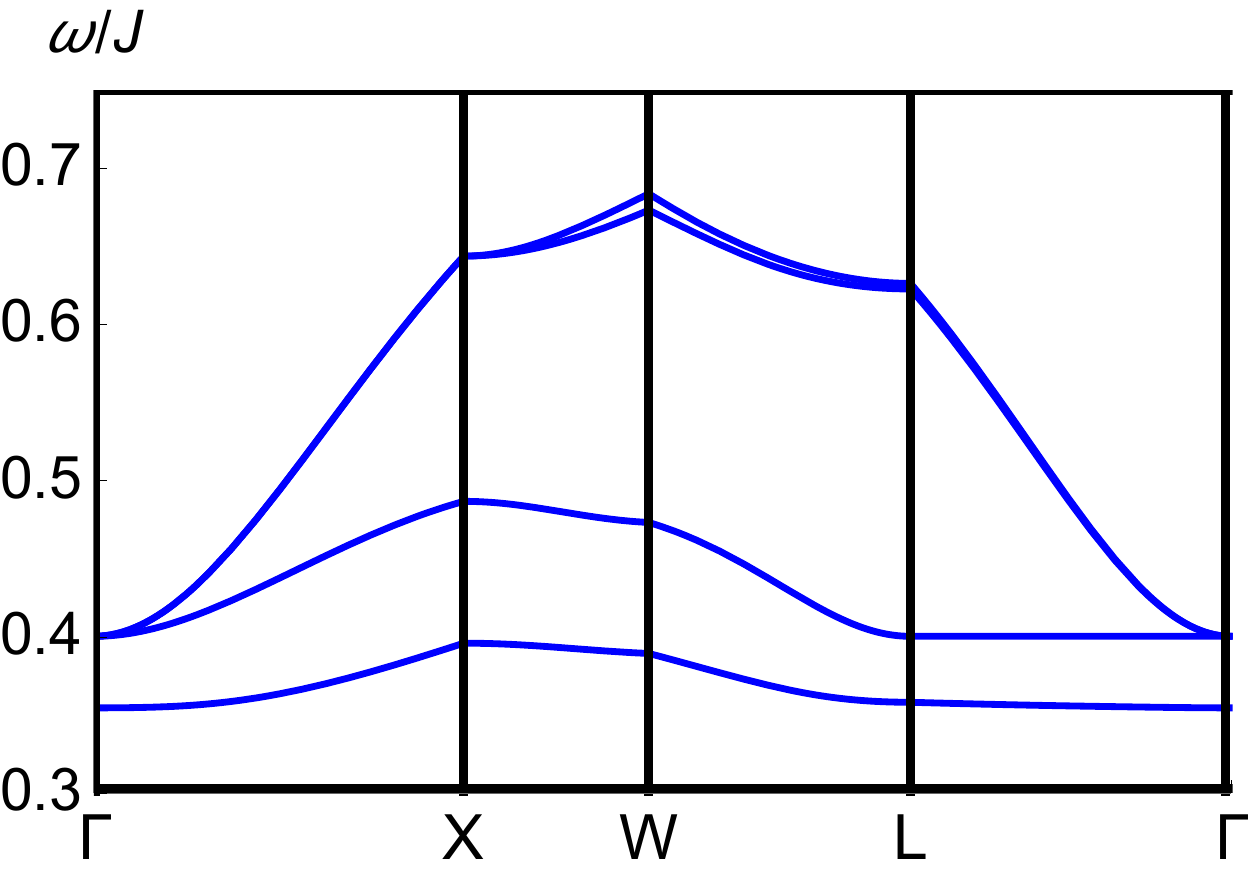}}~~~
\subfigure[]{\includegraphics[width=4.cm]{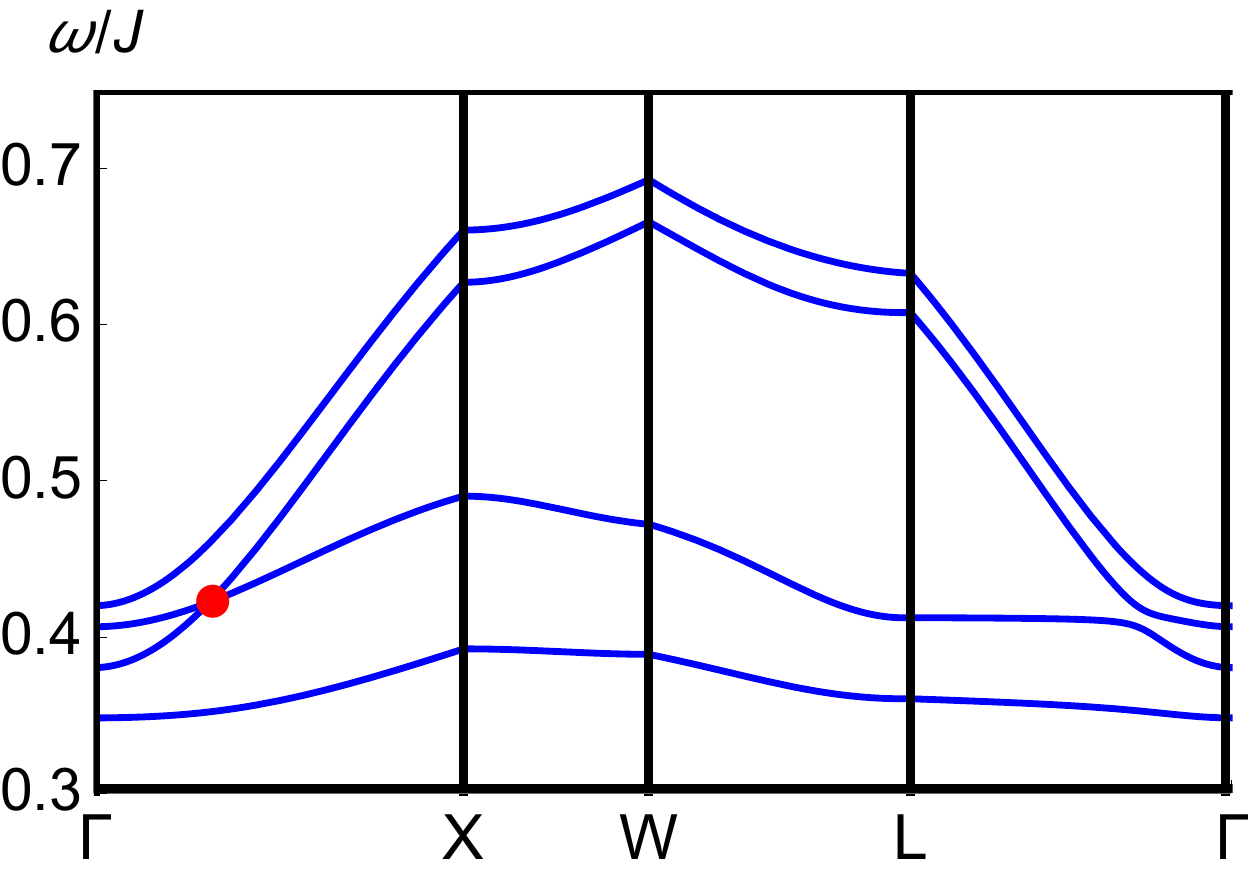}}
\caption{Spin-wav spectra with DM interaction. (a) The magnon bands along the path $\Gamma$-$X$-$W$-$L$-$\Gamma$ with $S=1/2, D=-0.2J, DM=0.18 J$. Note that the four bands split into a triple degeneracy and a singlet at $\Gamma$ point. (b) The magnon bands with applied magnetic field $B=0.05 J$ along (100) direction. The band-crossing point is indicated by a red point along $\Gamma$-$X$ line.}
\label{fig4}
\end{figure}

The ``direct'' (positive) DM interaction is found to favor the AIAO ground state~\cite{DM_pyrochlore,donnerer2016,supp}. Our spin-wave analysis for Eq.~\eqref{hamiltonian_DM} shows that the magnon bands at $\Gamma$ point also form a singlet and a three-dimensional $T_g$ representation as shown in Fig. \ref{fig4}(a), where $DM=0.18J$ \cite{donnerer2016}. Thus all the symmetry analysis can apply to such a case, namely, there will emerge two Weyl points in pyrochlore under magnetic field, and there will emerge four Weyl points in breathing pyrochlore under strains. Here, we explicitly show the emergent Weyl points in pyrochlore under magnetic field which is indicated by red point in Fig. \ref{fig4}(b), where the magnetic field $B=0.05J$ is along $(100)$ direction. 

\textit{Conclusions.---}With the aid of symmetry analysis, we show that under magnetic fields pyrochlores with AIAO orders can host Weyl magnons. Moreover, the pyrochlores exhibit nodal line upon a uniaxial strain. We also predict that four Weyl points would emerge in breathing pyrochlore antiferromagnets with AIAO orders upon uniaxial strains. To confirm the predictions, we perform a linear spin-wave calculation to a simple microscopic spin model. Owing to the ubiquitous existence of AIAO orders in nature and achievable technique in experiments, our findings shed light on experimental realization of Weyl magnons.

We remark on the possible experimental detections of these topological Weyl magnons. The bulk Weyl points can be detected by inelastic neutron scattering. While for the magnon arcs, for instance, there is a magnon arc connecting the projections of opposite-charged Weyl points in $(001)$ surface Brillouin zone of the pyrochlore under magnetic fields, it should be possible to detect them by using surface-sensitive probes, such as high-resolution electron energy loss spectroscopy, or helium atom energy loss spectroscopy. Besides the spectroscopic experiments, the Weyl magnon semimetals would also lead to anomalous thermal Hall effects \cite{li2013, ohe2013, mertig2014a, mertig2014b, owerre2016a, owerre2016b}, just like in the Weyl semimetals \cite{ran2011,burkov2012b}. In the materials showing the AIAO orders, such as the pyrochlore oxides Eu$_2$Ir$_2$O$_7$ \cite{sagayama2013, clancy2016}, and Cd$_2$Os$_2$O$_7$ \cite{yamaura2012}, the breathing pyrochlores Ba$_3$Yb$_2$Zn$_5$O$_{11}$ \cite{rau2016}, our predictions can be verified by above techniques experimentally.

\medskip

We would like to thank Hong Yao and Luyang Wang for helpful discussions. S.-K. Jian is supported in part by the NSFC under Grant No. 11474175 and by the MOST of China under Grant No. 2016YFA0301001. W. Nie is supported by the NSFC under Grant No. 
11704267 and by the MOST of China under Grant No. 2017YFB0405700. W. Nie acknowledge the hospitality of Renmin University of China and Tsinghua University, where part of the work was initiated.

\begin{widetext}

\section{Supplemental Material}
\renewcommand{\theequation}{S\arabic{equation}}
\setcounter{equation}{0}
\renewcommand{\thefigure}{S\arabic{figure}}
\setcounter{figure}{0}
\renewcommand{\thetable}{S\arabic{table}}
\setcounter{table}{0}

\subsection{A. The $\vec k \cdot \vec p$ theory}

The point group of pyrochlore lattice is $O_h$. The AIAO oder preserves $T_h$ group. The magnon bands have a triple degeneracy at $\Gamma$ point, which forms $T_{g}$ representation. Now we construct the $\vec k \cdot \vec p$ theory near the $\Gamma$ point. Three spin-one matrices serve as $T_{g}$ representations \cite{luttinger1956},
\bea
	L_x= \left( \ba{cccc} 0 & 0 & 0 \\
				0 & 0 & i \\
				0 & -i & 0  \ea\right),
	L_y= \left( \ba{cccc} 0 & 0 & -i \\
				0 & 0 & 0 \\
				i & 0 & 0  \ea\right),
	L_z= \left( \ba{cccc} 0 & i & 0 \\
				-i & 0 & 0 \\
				0 & 0 & 0  \ea\right).	
\eea
Using these matrices, one can construct all nine 3$\times$3 hermitian matrices: $I_3, L_x, c.p., \{L_x, L_y\}, c.p., 2L_z^2-L_x^2-L_y^2, L_x^2-L_y^2$, where $I_3$ is identity matrix and $c.p.$ means cyclic permutations. On the other hand, the momentum $p_i$ is a $T_{u}$ representation. We can construct the Hamiltonian by using these two representations, as shown in Table. S1. Then the most general Hamiltonian up to quadratic order in momentum is given by
\bea
	\mathcal{H}_T(\vec{p}) &=&  \alpha_1 |\vec{p}|^2 + \alpha_2 \sum_i p_i^2 L_i^2 + [p_x p_y( \alpha_3 \{ L_x, L_y \} + \alpha_4 L_z)+ c.p.],
\eea
where $\alpha_i$ are real parameters. Note that $\alpha_4$ term breaks time reversal symmetry. By fitting to the spin wave dispersion for $S=1/2, J=J'=1, D=-0.2$ in $10\%$ Brillouin zone, these parameters are given by $\alpha_1=-2.75 \times 10^{-3},\alpha_2=30.15 \times 10^{-3}, \alpha_3=26.03 \times 10^{-3}, \alpha_4=18.80 \times 10^{-3}$.

\begin{table}[h]
\centering
\caption{Representations of $T_h$ group constructed from $\vec L$ and $\vec p$.}
\begin{tabular}{|c|c|c|}
\hline
 ~~~~~Reps.~~~~~& ~~~~~~~$\vec L$~~~~~~~~~~~~ & ~~~~~~$\vec p$~~~~~~ \\
\hline
$A_g$ &1 & $p_x^2+p_y^2+p_z^2$     \\
\hline
$E_g$ & $(2L_z^2-L_x^2-L_y^2, L_x^2-L_y^2)$ & $(2p_z^2-p_x^2-p_y^2, p_x^2-p_y^2)$     \\
\hline
$T_g$ & $(L_x, L_y, L_z)$ and $(\{L_x, L_y\},\{L_y, L_z\},\{L_z, L_x\})$ & $(p_xp_y, p_yp_z, p_zp_x)$     \\
\hline
$T_u$ &  & $(p_x, p_y, p_z)$ \\
\hline
\end{tabular}
\end{table}

\subsection{B. Nodal-line magnons in strained pyrochlores}
To lower the symmetry of the system, one adds uniaxial strains along $(001)$ direction. The symmetry group is lowered to $D_{2h}$. To capture the strain, we add $\mathcal{H}_\text{strain}= \beta_\text{strain}(2L_z^2-L_x^2-L_y^2)$ into the Hamiltonian. There is a band crossing along $(100)$ [or $(010)$] direction though the double degeneracy is split. The crossing point is at $(\pm Q,0,0)$ and $(0, \pm Q,0)$, where $Q=\sqrt{3 \beta_\text{strain}/\alpha_2}$. Note that $\beta_\text{strain}/\alpha_2$ must be positive to have a band crossing. 
Actually, there emerges a nodal line at $p_z=0$ plane, given by the function
\bea
	6\beta_\text{strain}^2+4(\alpha_3^2+\alpha_4^2)p_x^2 p_y^2+\alpha_2^2 (p_x^2-p_y^2)^2-\alpha_2(p_x^2+p_y^2)=0.
\eea
The nodal-line is protected by $\sigma_h$ horizontal reflection symmetry. 

\subsection{C. Weyl magnons in strained breathing pyrochlores}
As shown in the main text, applying uniaxial strains to breathing pyrochlore with AIAO order leads to four Weyl points. Diagonalizing $\mathcal{H}_T+\mathcal{H}_\text{strain}+ \mathcal{H}_D$ leads to four Weyl points located at $(\pm Q_2,0,0)$ and $(0,\pm Q_2,0)$, where $Q_2= \sqrt{(3\alpha_2\beta_\text{strain}+ \beta_D^2)/\alpha_2^2}$.

\subsection{D. Spin wave analysis}

Since there is not $U(1)$ symmetry of Holstein--Primakoff bosons, we can write the Hamiltonian as $H= \sum_{\vec p} \Phi^\dag_{\vec p} \mathcal{H}(\vec p) \Phi_{\vec p}$ in spinor space $\Phi_{\vec p}= (a_{\vec p,\mu}, a_{-\vec p, \mu}^\dag )$, where
$\mathcal{H}(\vec p)= \Big( \ba{cccc} A(\vec p)/2 & B^\dag(\vec p) \\
	B(\vec p) & A^T(-\vec p)/2 \ea \Big).$
The local frame for the four sublattices of the pyrochlore lattices is shown in Table~\ref{table1}.
The expressions of the matrix elements of $A, B$ are given in the main text. The dispersions given in the main text are obtained by diagonalizing $g\mathcal{H}$, where $g=diag(I_{4\times4},-I_{4\times4})$, including properly $\delta H'_\text{strain}$ and (or) $\delta H_Z$.

In the strained breathing pyrochlore, applying an external magnetic fields shift the positions of two Weyl points as shown in the main text. Keep increasing the magnetic fields, there emerge four extra Weyl points which are shown in Fig. \ref{figs1}: a schematic plot of total six
Weyl points in Fig. \ref{figs1}(a) with the projections of directions of Berry curvatures in Fig. \ref{figs1}(b). Note that these points in Fig. \ref{figs1}(b) do not indicate the Weyl points locate at $p_x=0$ plane since there is no symmetry to fix it.

\begin{table}[b]
\centering
\caption{Local frame in each sublattice for AIAO orders, indicated in Fig. 1(b). \label{table1}}
\begin{tabular}{|c|c|c|c|}
\hline
 ~~~~$\mu$~~~~& ~~~~~~~~~$\hat x_\mu$~~~~~~~~~ & ~~~~~~~~~$\hat y_\mu$~~~~~~~~~ & ~~~~~~~~~$\hat z_\mu$~~~~~~~~~  \\
\hline
$0$ & $\frac{1}{\sqrt{2}}(-1,1,0)$ & $\frac{1}{\sqrt{6}}(-1,-1,2)$ & $\frac{1}{\sqrt{3}}(1,1,1)$    \\
\hline
$1$ & $\frac{1}{\sqrt{2}}(-1,-1,0)$ & $\frac{1}{\sqrt{6}}(-1,1,-2)$ & $\frac{1}{\sqrt{3}}(1,-1,-1)$    \\
\hline
$2$ & $\frac{1}{\sqrt{2}}(1,1,0)$ & $\frac{1}{\sqrt{6}}(1,-1,-2)$ & $\frac{1}{\sqrt{3}}(-1,1,-1) $   \\
\hline
$3$ & $\frac{1}{\sqrt{2}}(1,-1,0) $& $\frac{1}{\sqrt{6}}(1,1,2)$ & $\frac{1}{\sqrt{3}}(-1,-1,1)$    \\
\hline
\end{tabular}
\end{table}

\begin{figure}[t]
\subfigure[]{\includegraphics[width=5.cm]{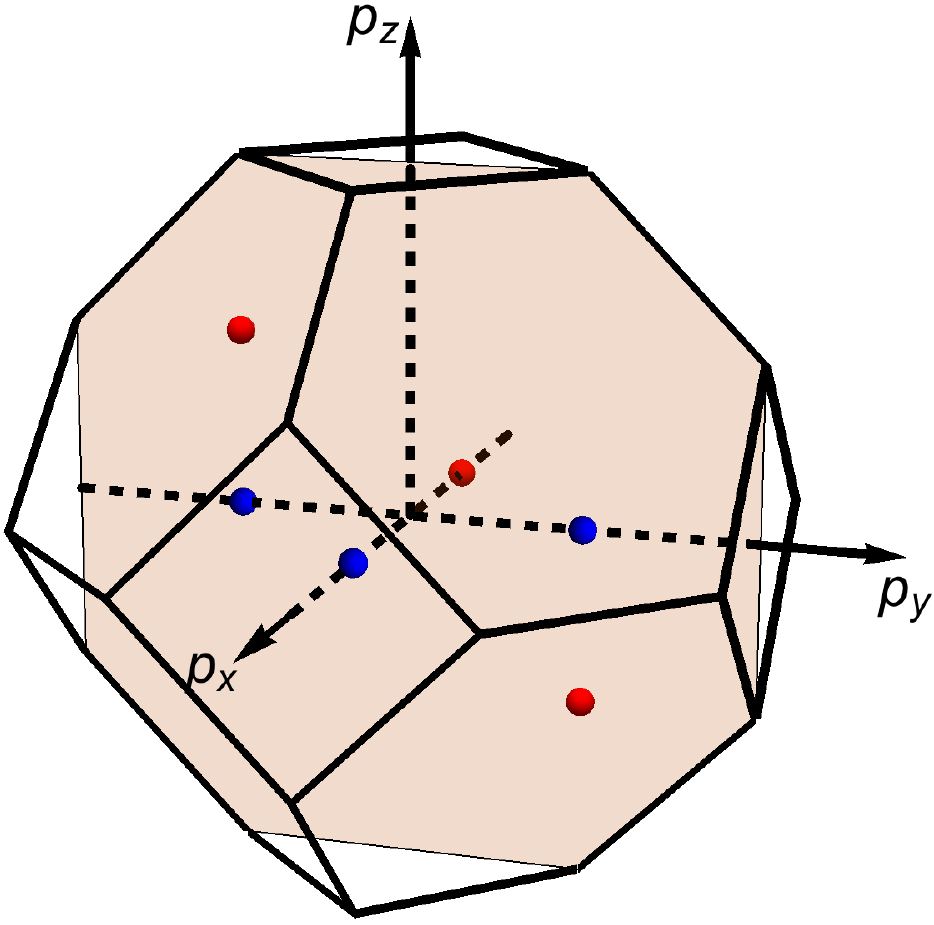}}~~~
\subfigure[]{\includegraphics[width=5.cm]{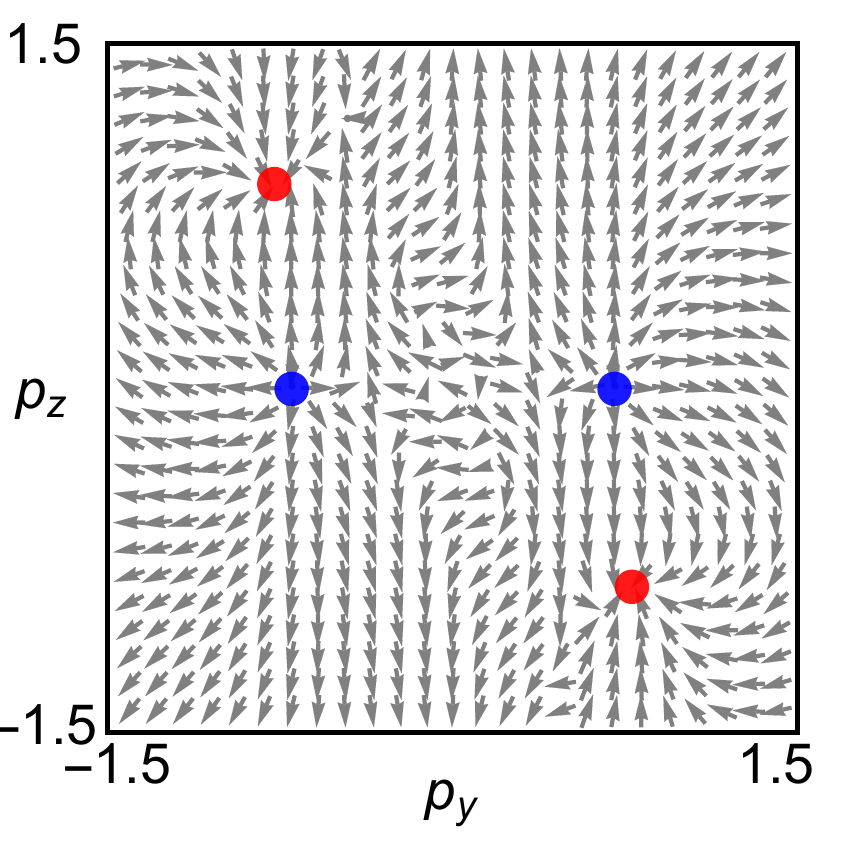}}
\caption{ (a) A schematic plot of six Weyl points in Brillouin zone. Two of them are restricted in $p_x$ axis due to $C_2$ symmetry along $(100)$ symmetry. (b) The projections of the directions of Berry curvatures in $p_x=0$ plane.}
\label{figs1}
\end{figure}

\subsection{E. Spin-wave analysis with Dzyaloshinskii-Moriya interaction}
The Dzyaloshinskii-Moriya (DM) vectors are given by \cite{DM_pyrochlore,Onose2010}
\bea
&& \vec{D}_{03}=\frac{DM}{\sqrt{2}}(-1,1,0), \vec{D}_{12}=\frac{DM}{\sqrt{2}}(-1,-1,0), \vec{D}_{01}=\frac{DM}{\sqrt{2}}(0,-1,1), \\
&& \vec{D}_{23}=\frac{DM}{\sqrt{2}}(0,-1,-1), \vec{D}_{13}=\frac{DM}{\sqrt{2}}(1,0,1), \vec{D}_{02}=\frac{DM}{\sqrt{2}}(1,0,-1).
\eea
Since AIAO is already the classical ground state for Eq.(1) in the main text, we only need to check AIAO is also a ground state for DM term. Assuming translation symmetry, the problem reduces to the configuration inside a unit cell which has only eight degrees of freedom corresponding to four spins. Expressed in local frame in Table \ref{table1}, one can do a variation with respect to these eight parameters. When $DM>0$, it turns out that the lowest energy is given by the configuration where each spin is parallel to the $\hat z_\mu$ direction in local frames, namely, the configuration corresponds to AIAO order.

\end{widetext}
\end{document}